\documentclass[12pt]{article}

\usepackage{amsmath}
\usepackage{amsfonts}

\newcommand{\aaa}{\mathcal{A}}
\newcommand{\bra}[1]{\pmb{\langle}#1\pmb{|}}

\newcommand{\ccc}{\mathcal{C}}
\newcommand{\cfield}{\mathbf{C}}
\newcommand{\dff}{\sc}
\newcommand{\eee}{\mathbf{e}}
\newcommand{\h}{\mathbf{H}}
\newcommand{\hhh}{\mathcal{H}}
\newcommand{\ia}{\Omega}
\newcommand{\idest}{{\itshape ie}}
\newcommand{\ket}[1]{\pmb{|}#1\pmb{\rangle}}
\newcommand{\ketbra}[2]{\ket{#1}\bra{#2}}
\newcommand{\rota}{\propto}

\begin{document}

\pagestyle{myheadings} \markboth{Superposed Quantum
Networks}{Superposed Quantum Networks}

\title{Superpositional Quantum Network Topologies}

\date{}

\author{Christopher Altman\thanks{Quantum Information Science and
Technology Project, ATIP, Tokyo, Japan, and Universiteit van
Amsterdam, The Netherlands, email {\texttt altmanc@admiral.umsl.edu}},
{Jaros\l{}aw Pykacz\thanks{Instytut Matematyki, Uniwersytet
Gda\'nski, Wita Stwosza 57, 80-952 Gda\'nsk, Poland, and Center
Leo Apostel of the Vrije Universiteit Brussels (VUB),
Krijgskundestraat 33, 1160 Brussel, email {\texttt
pykacz@delta.math.univ.gda.pl}}, Rom\`an R.
Zapatrin\thanks{Friedmann Lab. for Theoretical Physics, SPb UEF,
Griboyedova 30--32, 191023, St.Petersburg, Russia, email {\texttt
zapatrin@rusmuseum.ru}}}}

\maketitle

\begin{abstract}
\noindent We introduce superposition-based quantum networks
composed of (i) the classical perceptron model of multilayered,
feedforward neural networks and (ii) the algebraic model of
evolving reticular quantum structures as described in quantum
gravity. The main feature of this model is moving from particular
neural topologies to a quantum metastructure which embodies many
differing topological patterns. Using quantum parallelism,
training is possible on superpositions of different network
topologies. As a result, not only classical transition functions,
but also topology becomes a subject of training. The main feature
of our model is that particular neural networks, with different
topologies, are {\em quantum states}. We consider high-dimensional
dissipative quantum structures as candidates for implementation of
the model.
\end {abstract}


\medskip

\setlength{\unitlength}{.5mm}

\section*{Introduction}

Quantum learning networks have been suggested to offer new
domains for quantum algorithm design \cite{behrman, chrisley,
ventura}. Machine learning-inspired architectures are
self-organizing, robust, and ideal for such tasks as pattern
recognition and associative processing. Contemporary quantum
network models convey their advantage using superposed quantum
states on a fixed topological background. We suggest superposed
quantum {\em topologies} as a novel approach to quantum neural
networks, and present a model of learning and evolving
superposed quantum network topologies, or SQNTs. The
mathematical basis of this model is predicated upon two
existing formalisms:

\begin{itemize}
\item Neural networks: the classical learning model of
pattern recognition \cite{dorogov,rhw}
\item Algebraic quantum foam model: the formalism which
describes superposed and continuously evolving discrete
structures \cite{aphro,fas}
\end{itemize}

\noindent
In order to show how our model can be implemented we consider
multidimensional quantum systems admitting highly degenerate
states as described in \cite{savvidy} and the Bose-Einstein condensate
as outlined in \cite{vitiello}. \medskip

\noindent
An overview of the model is arranged as follows. We begin with a classical perceptron model, namely, that of a multilayered feedforward, weakly connected neural network. Then we extend this model, making it quantum by admitting the existence of superpositions of differing topological structures of neural networks using Rota algebraic formalism \cite{fas} as applied to describe the evolution of reticular patterns of quantum spacetimes. This means that

\begin{enumerate}
\medskip
\item We admit that the topological structure of a perceiving
entity can change
\item We admit that these changes may be continuous.
\end{enumerate}
\medskip

\noindent
At first sight these two requirements look contradictory: how
can discrete structures evolve continuously? These requirements
are reconciled in quantum mechanics. Along these lines we
present the model of superposition-based quantum network topologies.

\section{Interlude on reticular quantum spacetime
formalism}\label{s1}

Before coming to SQNT models in more detail and for the sake of
self-consistency, we recall the necessary constructions from
quantum mechanics used to describe reticular quantum spacetime.
The main feature of the quantum mechanical description of a
physical system is that we pass from configuration space to a
complex linear space, called the {\dff state space} of the
system. We shall consider systems with finite configuration
spaces, therefore their state spaces will be finite-dimensional
Hilbert spaces $\hhh=\cfield^N$.  Suppose we have a system
whose configuration space is $\{1,\ldots,n\}$. According to
quantum mechanics, it can be in pure (that is, dispersion free)
superposed state such as

\begin{equation}\label{expurestate}
\ket{\psi}
\,=\,
\cos\alpha\,\ket{1}
+
e^{i\phi}\sin\alpha\,\ket{2}
\end{equation}
\medskip

\noindent
Quantum observables, that is, measuring apparata are described
by self-adjoint operators in the state space of the system in
question. The values of the observables are the eigenvalues of
the operators; they are always real. Note that mutually
commuting (thus having the same eigenvectors) observables are
associated with the same measuring apparatus -- they are
interconverted just by relabelling the pointer's values. To
specify an observable $K$, we consider a decomposition of the
unit operator by orthogonal projectors and associate a {\em
number} with each projector. In Dirac notation this reads:

\begin{equation}\label{ediracop}
K\,=\,\sum_{i}k_i\:\ketbra{i}{i}
\end{equation}

\paragraph{Spatialization.} The {\em spatialization procedure} was developed in \cite{aphro} for the purpose of describing spacetime foam. Its primary feature is to associate discrete structures, rather than numerical values \eqref{ediracop}, with subspaces of the state space. More specifically, in standard quantum mechanics the state space is defined as a Hilbert space -- that is, a complex linear space with an inner product, $\h\times\h\to\cfield$. Each subspace of $\h$ can be associated only with its dimension as an integer. We associate with each subspace $\ccc$ a disjoint graph, whose number of vertices equals its dimension, $\dim$ $\ccc$. \medskip

\noindent The spatialization procedure links some of the vertices, thus associating the subspace $\ccc$ with richer structure than a cardinal number. The only requirement is for $\hhh$ to be endowed with an associative product structure, rather than an inner product. In some cases this structure already exists -- for instance, when $\hhh$ is a tensor
product of two copies of the same state space $\h$, that is
$\hhh=\h\otimes\h$. The associative product is defined on
factorable vectors as

\[
\phi\otimes\phi' \;\cdot\; \psi\otimes\psi' \;=\; <\phi',\psi>
\;\cdot\; \phi\otimes\psi'
\]

\medskip \noindent and then extended by linearity (this is nothing but a
usual matrix product). The spatialization procedure is briefly
outlined as follows: with any directed acyclic transitive graph
$G$ with $N$ vertices its {\dff Rota algebra} is associated
$\aaa$, whose elements are $N\times{}N$ matrices of the following
form. They have zero entries $a_{ik}$ when the vertices $i,k$ are
{\em not} connected in the graph $G$. For instance, consider the
graphs

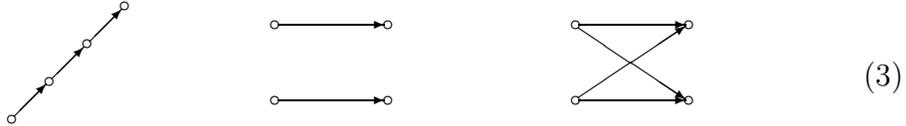
\begin{equation}\label{ecircgr}
\begin{picture}(200,25)(0,9)
\put(0,0){\circle{2}} \put(10,10){\circle{2}}
\put(20,20){\circle{2}} \put(30,30){\circle{2}}
\put(0.75,0.75){\vector(1,1){8.5}}
\put(10.75,10.75){\vector(1,1){8.5}}
\put(20.75,20.75){\vector(1,1){8.5}}
\put(70,5){\circle{2}}
\put(70.75,5){\vector(1,0){28.5}}
\put(100,5){\circle{2}}
\put(70,25){\circle{2}}
\put(70.75,25){\vector(1,0){28.5}}
\put(100,25){\circle{2}}
\put(150,5){\circle{2}}
\put(150.75,5){\vector(1,0){28.5}}
\put(150.75,5.75){\vector(3,2){28.5}}
\put(180,5){\circle{2}}
\put(150,25){\circle{2}}
\put(150.75,25){\vector(1,0){28.5}}
\put(150.75,24.25){\vector(3,-2){28.5}}
\put(180,25){\circle{2}}
\end{picture}
\end{equation}

\bigskip

\noindent For these examples, the appropriate Rota algebras take the following form

\medskip

\begin{equation}\label{ecircle}
\begin{array}{c@{\quad}c@{\quad}c}
\left(\begin{array}{cccc}
\ast & \ast & \ast  & \ast  \cr
0 & \ast & \ast  & \ast  \cr
0 & 0 & \ast & \ast \cr
0 & 0 & 0 & \ast \cr
\end{array}\right)
&
\left(\begin{array}{cccc}
\ast & 0 & \ast  & 0  \cr
0 & \ast & 0  & \ast  \cr
0 & 0 & \ast & 0 \cr
0 & 0 & 0 & \ast \cr
\end{array}\right)
&
\left(\begin{array}{cccc}
\ast & 0 & \ast  & \ast  \cr
0 & \ast & \ast  & \ast  \cr
0 & 0 & \ast & 0 \cr
0 & 0 & 0 & \ast \cr
\end{array}\right)
\end{array}
\end{equation}
\medskip

\noindent For a detailed account of Rota algebras and the
spatialization formalism supplied with examples the reader is referred to
Appendices A, B and \cite{aphro,fas}.

\paragraph{Neural metastructures.} The main feature of the
proposal put forward in this paper is to shift from a
particular structure of neural network as perceptron to something
different: a metasystem whose {\em states} are perceptrons.
This is accomplished as follows. Superposed quantum neural
networks are themselves linear spaces rather than
neural networks. With each subspace of the state space of
SQNT -- which is its quantum property, see \eqref{ediracop}
above -- a particular neural network configuration is associated
using the above mentioned spatialization procedure. In section
\ref{sqtain} we suggest to use the quantum features of SQNTs to
enhance the performance of the main task in neural networks,
namely, their training. \medskip

\noindent Realistic physical candidates to implement SQNTs have
already been proposed. The first is based upon the Bose-Einstein
condensate, on which a quantum brain model has been suggested by
G. Vitiello \cite{vitiello}. A second model has been suggested by
G. Savvidy \cite{savvidy}. Both have the crucial property needed
for our purposes, namely they have many individual degrees of
freedom and admit superpositions. We emphasize that no {\em ab
initio} association of states with graphs is needed, as graphs are
automatically produced as a consequence of the spatialization procedure.
SQNT implementation can also be modelled classically: as demonstrated in
\cite{werbos}, a broad class of quantum algorithms can be
simulated on classical systems.

\section{Neural networks and Rota algebras}

One of the basic tasks of neural networks is to function as perceptrons, 
that is to recognize signals for which we have 
no structural theory -- for instance, to recognize visual patterns.
In this section we review the basic principles on which
perceptrons are based, their learning and training in classical setting
and show how Rota algebras introduced above emerge in their description. We shall deal with multilayered feedforward NNs, such as

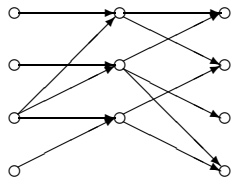
\[
\unitlength0.7mm
\begin{picture}(100,40)
\multiput(0,0)(0,10){4}{\circle{2}} 
\multiput(20,10)(0,10){3}{\circle{2}} 
\multiput(40,0)(0,10){4}{\circle{2}} 
\multiput(0.75,10)(0,10){3}{\vector(1,0){18.5}} 
\multiput(0.75,0.75)(0,10){2}{\vector(2,1){18.5}} 
\put(0.75,10.75){\vector(1,1){18.5}} 
\multiput(20.75,9.25)(0,10){3}{\vector(2,-1){18.5}} 
\put(20.75,19.35){\vector(1,-1){18.5}} 
\multiput(20.75,10.75)(0,10){2}{\vector(2,1){18.5}} 
\put(20.75,30){\vector(1,0){18.5}} 
\end{picture}
\]

\medskip
\noindent that is, their nodes can be arranged in layers so that (i) no nodes 
in a given layer communicate and (ii) the signals propagate only consecutively via layers. The state of a given neural network is the set of values, numerical or vectorial, assigned to its nodes. A time-step propagation changes the perceptron's state so that the value of a node $i$ at time $t+1$ depends on its own value and the values of nodes adjacent to $i$ at time $t$. 

\medskip
\noindent Besides their multilayered, high-dimensional structure the filtering and recognizing power of neural networks is based on the nonlinearity of their transition functions. In some cases, in particular while dealing with stability or elasticity issues, we may use linear
approximations. This is exactly the case we explore in this
paper: optimal topological configurations are assumed to be
robust under small permutations \cite{dorogov}. 

\paragraph{Training.} Initially one starts with a set of
patterns for which the classification is known.  Usually by
means of heuristic methods, the topological structure of the
network is chosen and then trained via input of known patterns
and subsequent adjustment of network parameters (transition
functions). Output signals are correlated with patterns from
different classes to be well-separated with respect to certain
criterion.  The most popular method to adjust transition
functions is error backpropagation (see, for instance,
\cite{rhw}).

\paragraph{Performance.} Signal propagation in the linear approximation can be viewed as a matrix multiplication which reduces to a number of
arithmetic operations. The more links there are between neurons,
the more computational resources are consumed by the process of
pattern recognition. In order for a neural network to be faster,
we should seek for sparser configurations.\medskip

\noindent
Therefore, the criterion for `good matching' should also take
performance into consideration. As an example we consider the
approach of {\em fast neural networks}, which are weakly
connected multilayered feedforward neural networks \cite{dorogov}.
The idea of these networks grows from
the FFT -- fast Fourier transform. When finding optimal
configurations of such networks, we restrict the allowed links
between neurons and force networks to be sparse. Note that in
our setting this requirement is followed automatically as we
are confined by the dimensionality of the state space $\hhh$.

\section{Superposition-based training}\label{sqtain}

The basic proposal of this paper is to replace training of
particular neural network configurations with training of
superposed ones. This can be achieved in the following steps:

\medskip

\noindent (i) SQNT is treated as a system described by quantum
mechanical formalism, in our special case represented by a matrix
algebra $\aaa$. Note that the term `quantum' here is a mere
indication of the rules of behavior of the object on which the SQNT
is implemented, whatever be its `real' nature.\medskip

\noindent (ii) Input signals are vectors from a representation
space of the algebra $\aaa$, as are output signals. Connection weights are represented by the matrix elements, and can further be represented by matrices of linear operators consequent of the spatialization procedure. \medskip

\noindent (iii) Learning is a matching algorithm which selects
a subset $\ccc$ of $\aaa$ for which the criterion of matching
between input and output signals is optimal. At this step we
obtain the optimal configuration on the metalevel.\medskip

\noindent (iv) In order to pass to a particular neural network
structure, we consider a subalgebra (rather than a subset) of
$\aaa$ which is the nearest to the subset $\ccc$.

\medskip

\noindent
Now let us dwell on the above issues. For the first step, an
appropriate physical system for SQNT to be implemented should
be found. The main requirement which it has to satisfy is to
possess sufficiently many degrees of freedom with controlled
access to them. The initial candidates for these purposes are
entangled quantum registers on which quantum computers
are based. An alternative model can be considered using the Bose-Einstein condensate
\cite{vitiello}.

\medskip

\noindent
So, we start with a quantum system $S$ with sufficiently many
degrees of freedom. To implement (ii), we must draw the
distinction between the classes of states treated as
states of the input register and between the states responsible
for the operations of multiplication. This step does not affect
the overall performance of the future perceptron, as it has to be
done only once (this is analogous to building a computer). The
work of the SQNT is to evolve according to a prescribed evolution,
starting from the initial state (= a pattern to recognize) to
the final one (= its classification identifier).\medskip

\noindent
An algorithm seeking for best matching (iii) can be realized as
follows.  We feed in a pattern from a given sample set (= given
prescribed state) and then let it evolve. The evolution is set in
such a way that it realizes the multiplication operation. To be
more precise, we consider any available evolution $U_t$ and then
label the states of SQNT in such a way that for the unit time
$\delta{}t$ the resulting action of $U_{\delta{}t}$ would be
multiplication of appropriate elements. This is possible due to
the arbitrary nature of labeling of eigenstates of operators (see
section \ref{s1}). \medskip

\noindent
The condition for learning to be successful is that the input
vector from a sample set should be unchanged. In physical terms,
this means patterns from the sample set should be
eigenvectors for the energy operator responsible for the
evolution. Why do we require it? The point is that we would like
our output signals to be well separated -- that is, they should be
stable with respect to small variations of the input vector. This,
in turn, requires the states of the overall register {\em (input,
output)} to be orthogonal. If they are eigenstates of the
generator of the evolution, this holds automatically. We denote by
$\ccc$ the appropriate eigenspace and call it {\dff calculation
subspace}.

\medskip

\noindent
The last item (iv) is a kind of `classical correspondence
principle' in quantum mechanics. SQNT operates as it
is -- namely, it undergoes quantum evolution -- but if we would like
to represent (perhaps with some losses) its work in terms of
neural networks, we proceed in the following way. We deal with
the calculation subspace $\ccc$ of our state space, which we
obtained as a result of the best matching algorithm, and we have the
evolution operator in our disposal which is interpreted as
multiplication.\medskip

\noindent
The spatialization procedure can be applicable
to $\ccc$ only if it is closed with respect to the
multiplication, while this requirement may not hold for
eigenspaces of the energy operator. That is why, following the
lines of quantum mechanics, we consider the nearest {\em
subalgebra} to $\ccc$ and immediately interpret it as a neural
network. This suggestion is in full accordance with quantum
measurement theory, where the wavefunction of the system immediately
`collapses' to an eigenstate of the appropriate operator.

\section{Concluding remarks}

\noindent Presently only a handful of quantum algorithms exist,
and these are confined to a limited set of specialized
applications. Quantum learning architectures offer the potential
to expand this domain to a much broader class of functionality. We
have outlined the fundamentals of a quantum fast-training pattern
recognition model -- {\em superpositional quantum network
topologies}, or SQNT -- which provides a rich source for the development of a novel class of quantum algorithms. The distinguishing feature of our model lies in its ability to utilize coherent superposition of unique
topological configurations of neural networks.
\medskip 

\noindent We suggest two ways to implement our model in physical media based on high-dimensional dissipative quantum systems \cite{savvidy,vitiello}. Quantum simulation methods offer an immediate candidate for study of the model: non-classical dynamics can be efficiently simulated on
ensembles of states of classical Turing machines ruled by
second-order differential equations, see \cite{werbos} for more
detail. It should be mentioned that although macroscopic physical systems
are comprised of quantum components, quantum phenomena in the
macroscopic realm are usually invisible due to averaging over a
large number of degrees of freedom. Nevertheless, the existence of
macroscopic systems exhibiting quantum properties cannot
be neglected in many subdomains of the life sciences, especially
in molecular biology \cite{aertsgrib}. For example, A. S.
Davydov in his book `Biology and Quantum Mechanics' \cite{davydov}
studied collective soliton excitations in large protein molecules and applied them to such biological phenomena as membrane transport, nerve impulse conduction, and muscle contraction. Extending beyond relevance to the current model, we suggest this could be a potential explanation of the
surprisingly high adaptation capabilities of living organisms.
Classical models of learning and response to continuously changing environments require immense computational resources. We interpret this as a signal that perhaps quantum models could be more efficient for this purpose.

\paragraph{Acknowledgments.} The authors highly appreciate
the attention to their work offered by A. Dorogov, S. Krasnikov
and P. Werbos. This work was carried out within the ATIP Quantum
Information Science and Technology Project, the project G.0335.02
of the Flemish Fund for Scientific Research and the research
program {\em`Quantum Theory in the Early Universe,'} under the
auspices of the Russian Basic Research Foundation (RFFI).

\newpage

\section*{Appendix A: Rota algebras of transitive graphs}\label{ssrotalg} 

\paragraph{Rota algebras in Dirac notation.} Let $X$ be a reflexive transitive graph. 
For brevity a pair of nodes $i,j$ of $X$ connected by an arrow $i\to j$ is said to be {\dff tending}. 
Consider the linear space $\ia$ whose basis 
$\ketbra{i}{j}$ is labelled by tending pairs $i\to j$ of nodes of $X$. 

\begin{equation}\label{edefrotalg}
\ia(X)=
\left\{
\sum_{i,j\in X}\limits\ketbra{i}{j}
\quad\mbox{such that}\quad i\to j
\right\}
\end{equation}

\noindent In the sequel, when no confusion occurs, we omit the 
notation of the graph $X$ in parentheses and simply 
write

\[
\ia=\ia(X)
\]
 
\noindent Define the product on $\ia$ by setting it on its basic 
elements:

\begin{equation}\label{edef22}
\ketbra{i}{j}\ketbra{k}{l}\quad =\quad
\left\lbrace\begin{array}{lcl}
\ketbra{i}{l} &,& \hbox{if} \quad j=k \cr
0 &,& \hbox{otherwise}
\end{array}\right.
\end{equation}

\noindent Note that $\ketbra{i}{l}$ in \eqref{edef22} is always 
well-defined since the graph $X$ is assumed to be 
transitive, that is why the existence of darts $i\to j$ and $j\to k$ 
always enables the existence of $i\to k$. 
The space $\ia$ with the product \eqref{edef22} is 
called the {\dff Rota algebra} of the topological space $(X,\to)$. 
These algebras were first introduced in \cite{rota} in the context 
of combinatorial theory. 

\paragraph{The matrix representation of Rota algebras.} Given the 
Rota algebra of a transitive graph $X$, its standard matrix 
representation is obtained by choosing the basis of $\ia$ 
consisting of the elements of the form $\ketbra{i}{k}=\eee_{ik}$, 
with $ik$ ranging over all tending pairs $i\to k$ of nodes of 
$X$. The matrices $\eee_{ik}$ (called matrix units) are 
defined as follows:

\begin{equation}\label{eab}
\eee_{ik}(m,n)= 
\left\lbrace\begin{array}{rl}
1 & \mbox{$m=i$ and $n=k$ (provided $i\to k$)}\cr
0 & \mbox{otherwise}
\end{array}\right.
\end{equation}

\noindent where $\eee_{ik}(m,n)$ stands for the element in the 
$m$-th row and the $n$-th column of the matrix $\eee_{ik}$. We can 
also extend the ranging to {\em all} pairs of elements of $X$ by 
putting $\eee_{ik}\equiv 0$ for $i\not\to k$. Then the product 
\eqref{edef22} reads:

\begin{equation}\label{mprod}
\eee_{ik}\eee_{i'k'} = \delta_{ki'}\eee_{ik'}
\end{equation}

\medskip 

To specify a Rota algebra in the standard matrix representation we 
fix the template matrix replacing the unit entries in the incidence 
matrix $I_{ik}$ of the graph $X$: 

\[ 
I_{ik} = 
\left\lbrace\begin{array}{rl}
1 & i\to k\cr
0 & \mbox{otherwise}
\end{array}\right.
\]

\noindent by wildcards $*$ ranging independently over all numbers.  
For instance, the algebra associated with the two-node graph

\(\unitlength=0.2ex
\begin{picture}(23,4)
\put(0,4){\circle*{3}}
\put(20,4){\circle*{3}}
\put(3,4){\vector(1,0){15}}
\end{picture}
\) 
has the the following template matrix 

\[
\ia({\mbox{\(\unitlength=0.1ex
\begin{picture}(23,4)
\put(0,4){\circle*{3}}
\put(20,4){\circle*{3}}
\put(3,4){\vector(1,0){15}}
\end{picture}
\)}})= 
\left( \begin{array}{cc}
* & * \\
0 & *
\end{array} \right)
\,=\,
\left\lbrace \left.
\left( \begin{array}{cc}
a & b \\
0 & c
\end{array} \right)
\right\vert
\;
a,b,c \in \cfield
\right\rbrace
\]

\medskip 

So, we see that any transitive graph can be described in 
terms of a finite-dimensional algebra (for further details we refer to 
\cite{fas}).

\section*{Appendix B: Spatialization procedure}\label{ssrotatop}

Here we describe the spatialization procedure which associates a graph with an 
arbitrary finite-dimensional algebra. 

\paragraph{The emergence of nodes.} Let us start with a given 
finite-dimensional associative (and non-commutative, in general) 
algebra $\ia$. According to standard conceptions and methods of 
modern algebraic geometry, as well as the general algebraic 
approach to physics, we introduce the points of $\ia$ as its irreducible 
representations (IRs). So, the first step of the specialization 
procedure is creating (or finding) points of $\ia$ (which will become 
nodes of the future graph):

\begin{equation}\label{eptsirs}
\{\,\mbox{points}\,\} =
\{\,\mbox{IRs}\,\}
\end{equation}

\paragraph{Standard set with nonstandard topology.} When the first 
spatialization step in a standard way \eqref{eptsirs} is done we may 
wish to proceed by connecting the set of nodes by arrows.  
This problem is mathematically equivalent to equipping the set of points of the 
algebra by a topology. There are standard recipes for this step as well like, 
say, the Zariski topology on the prime spectrum of $\ia$. Unfortunately, on 
finite-dimensional algebras this topology is always discrete, which 
leaves us no chance to fit the above requirement of being 
non-Hausdorff ({\idest}, not $T_{2}$). In terms of graphs that means that the standard recipes can not help us to create arrows. So, we are compelled 
to find another topology. For these purposes the Rota topology is suggested 
(first it was introduced in \cite{fas}). 

\medskip 

Let $\ia$ be a finite-dimensional algebra. Denote by $X$ the set of 
points of $\ia$, each of which we shall associate with a prime 
ideal in $\ia$. Consider two points (representations of $\ia$)  
$i,j\in X$ and denote by $\ker{i},\ker{j}$ their kernels. 
Both of them, being kernels of representations, are two-sided ideals in $\ia$, in particular,  subsets of $\ia$, hence both of the following expressions make 
sense:

\[
\ker{i}\cap \ker{j} \subset \ia 
\qquad\mbox{and}\qquad
\ker{i}\cdot \ker{j} \subset \ia
\] 

\noindent the latter denoting the product of subsets of $\ia$: 
$\ker{i}\cdot \ker{j}=\{a\in \ia\mid\, \exists u\in \ker{i},\, v\in \ker{j}:\, uv=a\}$. 
Since $\ker{i},\ker{j}$ are ideals, we always have the inclusion $\ker{i}\cdot 
\ker{j}\subseteq \ker{i}\cap \ker{j}$ which may be strict or not. Define the relation 
$\rota$ on $X$ as follows: 

\begin{equation}\label{edefrota}
i\rota j\quad
\mbox{if and only if} \quad
\ker{i}\,\ker{j}\neq \ker{i}\cap \ker{j}
\end{equation}

\medskip 

Then the {\dff Rota topology} is the weakest one in which $i\rota j$ 
implies the convergence $i\to j$ of the point $i$ to the point 
$j$.  Explicitly, the necessary and sufficient conditions for $i$ 
to converge to $j$ in the Rota topology reads: 

\begin{equation}\label{eqxy}
i\to j
\quad\mbox{if and only if}\quad
\exists k_0,\ldots,k_t,\ldots,k_n\,\mid\,
k_0=i,\,k_n=j;\,k_{t-1}\rota k_t
\end{equation}

\noindent This operation is called the transitive closure of the 
relation $\rota$. Note that, in general, the Rota topology can be 
defined on any set of ideals.

\medskip 

It was proved by Stanley \cite{stanley} that in that particular 
case when $\ia$ is the Rota algebra of a reflexive transitive graph, 
then its spatialization $X$ endowed with the Rota topology 
\eqref{eqxy} is homeomorphic to the initial topological space. 
However, in general if we have two reflexive transitive graphs and an 
arrow-preserving mapping between them, their Rota algebras may {\em not} 
be homomorphic. Recently, `good' classes of reflexive transitive graphs 
topological spaces were discovered for which the transition to Rota algebras 
is functorial \cite{dgreech,iasc}. In our approach, it supports the existence of 
the classical limit.

\end{document}